\documentclass[%
superscriptaddress,
preprint,
amsmath,amssymb,
aps,
onecolumn
]{revtex4-1}

\usepackage{graphics} 
\usepackage{epsfig} 
\usepackage{amsmath} 
\usepackage{amssymb}  
\usepackage{multirow} 
\usepackage[table]{xcolor} 

\definecolor{Gray}{gray}{0.85}
\definecolor{LightCyan}{rgb}{0.88,1,1}

\usepackage{color}
\usepackage{url}

\begin{document}

\title{
An Agent-Based Model for Bovine Viral Diarrhea
}

\author{Jason Bassett}
\email{j.bassett@tu-berlin.de}
\affiliation{Institut f\"ur Theoretische Physik, Technische Universit\"at Berlin, Hardenbergstra{\ss}e 36, 10623 Berlin, Germany}

\author{Pascal Blunk}
\affiliation{Institut f\"ur Theoretische Physik, Technische Universit\"at Berlin, Hardenbergstra{\ss}e 36, 10623 Berlin, Germany}
\affiliation{calimoto GmbH Babelsberger Stra{\ss}e 12, 14473 Potsdam, Germany}

\author{Thomas Isele}
\affiliation{Institut f\"ur Theoretische Physik, Technische Universit\"at Berlin, Hardenbergstra{\ss}e 36, 10623 Berlin, Germany}

\author{J\"orn Gethmann}
\email{joern.gethmann@fli.de}
\affiliation{Institute of Epidemiology, Friedrich-Loeffler-Institut, S\"udufer 10, 17493 Greifswald - Insel Riems, Germany}


\author{Philipp H\"ovel}
\email{philipp.hoevel@ucc.ie}
\affiliation{Institut f\"ur Theoretische Physik, Technische Universit\"at Berlin, Hardenbergstra{\ss}e 36, 10623 Berlin, Germany}
\affiliation{School of Mathematical Science, University College Cork, Cork T12 XF64, Ireland}

\date{\today}

\begin{abstract}
We present an exhaustive description of a stochastic, event-driven, hierarchical agent-based model designed to reproduce the infectious state of the cattle
disease called Bovine Viral Diarrhea, for which the livestock-trade network is the main route of spreading.
For the farm-node dynamics, it takes into account a vast number of breeding, infectious and animal movement mechanisms via a susceptible-
infected-recovered type of dynamics with an additional permanently infectious class. The interaction between the farms is described by a supply-demand farm manager mechanism governing the network structure and dynamics. The model includes realistic disease and breeding dynamics and allows to study numerous mitigation strategies of present and past government regulations, including different testing and vaccination scenarios.
\end{abstract}

\maketitle

\section{Model Description}
In the paper, we present and describe thoroughly the agent-based model following the \emph{Overview, Design concepts and Details} (ODD) protocol after its last revision in \cite{GRI10} and after the example of \cite{THU17}.
The source code of the model can be found in the repository  \url{https://github.com/Yperidis/bvd_agent_based_model}. The proposed approach consists of a hierarchical event-driven, stochastic agent-based model written in \textsc{C++} for single-thread execution, which describes the spread of \textit{Bovine Viral Diarrhea} (BVD) between animals
in the herd, the farm (through contact) and the in between farm level (through animal movements).

The whole simulation is designed to accommodate the particularities pertinent to the agricultural cattle system of Germany. In addition, all the results and the related sensitivity analysis performed were generated using farm size distributions of federal German states as extracted from the \textit{Herkunftssicherungs- und Informationssystem f\"{u}r Tiere} (HI-Tier) database (\url{https://www.hi-tier.de}). The simulation is to serve as a tool for approximating the current status of the BVD dynamics within the German agricultural system and most importantly for assessing the effect of various considered strategies implemented as policies by the administration in a cost effective manner to promote the eradication of BVD.

\section{Entities, State Variables and Scales}\label{sec:St_vars}

There are four hierarchical levels of the simulation, \textit{the System}, \textit{the Farm}, \textit{the Herd} and \textit{the Animal}, with the \textit{System} being the superset of all and the \textit{Animal} being the smallest unit. As will be apparent in the upcoming lists the animal agent entity is by far the most complex of all the levels, as it contains events and variables to model breeding, infectious, movement and trading, testing and vaccinating features.

\section{Details on the Numerical Implementation}
\begin{enumerate}
	\item \textbf{Pregnancy Related Cow Events}\\
	Note that breeding and health related states are intertwined in this category.
	\begin{itemize}
		\item \textit{No calf}, i.e. a non-pregnant female cattle.
		\item \textit{Susceptible} cattle.
		\item \textit{Persistently infected} cattle.
		\item \textit{Immune}, i.e. cattle with life-long immunity.
		\item \textit{Cripple}, i.e. a cow or heifer which gives birth to a malformed calf, which is to be put down immediately. 
		\item \textit{Abort}, i.e. a cow or heiferwhich is going to have an abortion.
		\item \textit{Infertile}, i.e. a cattle which has met the criteria to be classified as such.
	\end{itemize}
	\item \textbf{Infectious States}
	\begin{itemize}
		\item \textit{Susceptible} (S)
		\item \textit{Transiently infected} (TI)
		\item \textit{Immune} (R)
		\item \textit{Persistently infected} (PI)
	\end{itemize}
	\item \textbf{Test Related States}
	\begin{itemize}
		\item \textit{No status}, i.e. the animal has not been tested.
		\item \textit{Negative} test
		\item \textit{Positive} test
		\item \textit{Positive once}, i.e. an animal which has been tested positive once.
		\item \textit{Positive twice}, i.e. an animal which has been tested positive twice.
		\item \textit{Positive cow}, i.e. an animal which has been tested positive and has had offspring prior to testing.
	\end{itemize}
	\item \textbf{Animal Trade Related Criteria}
	\begin{itemize}
		\item \textit{Calf} (female).
		\item \textit{Heifer pre breeding}, i.e. a heifer which is not ready to be inseminated.
		\item \textit{Heifer ready for breeding}
		\item \textit{Infertile}
		\item \textit{Pregnant}
		\item \textit{Dairy cow}
		\item \textit{Old cow}
		\item \textit{Male calf}
		\item \textit{Young bull}
		\item \textit{Old bull}
		\item \textit{Number of types}. Accounting for future extensions (additional criteria).
	\end{itemize}
	\item \textbf{Events}\\
	Given in descending priority. Each one refers to a single hierarchical level of the system, i.e. to the system level, to the herd level, to the farm level or to the animal level.
	\begin{itemize}
		\item \textit{Change containment strategy}. System level.
		\item \textit{Jungtier small group}. Farm level. \textit{Jungtier} refers to the young calf window strategy.
		\item \textit{Jungtier exec}. System level (triggers \textit{Jungtier small group}).
		\item \textit{Quarantine end}. Farm level.
		\item \textit{Virus test}. Refers to a blood, antigen (virus) test initiated from the young calf window strategy. Animal level.
		\item \textit{Antibody test}. Animal level. Refers to a serological test (blood).
		\item \textit{Test}, accounting for the virus test through tissue testing (ear tag). Initiated via birth. Animal level.
		\item \textit{Manage}, calls the farm manager. System level.
		\item \textit{Stop}, halts the simulation. System level.
		\item \textit{Write output}, writes the specified output to a file. System level.
		\item \textit{Log output}, writes the specified output to the memory. System level.
		\item \textit{Abortion}. Animal level.
		\item \textit{Insemination}. Animal level.
		\item \textit{Conception}. Animal level.
		\item \textit{Birth}. Animal level.
		\item \textit{Death}. System level.
		\item \textit{End of MA}, signifies the expiry of the maternal antibody effect for a calf. Animal level.
		\item \textit{Infection}. Animal level.
		\item \textit{Recovery}. Animal level.
		\item \textit{Slaughter}. System event.
		\item \textit{Culling}, accounting for extensions of \textit{Slaughter}. System event.
		\item \textit{Vaccinate}. Animal level.
		\item \textit{End of vaccination}, signifies the expiry of a vaccination's effect. Animal level.			
		\item \textit{Trade}. Farm level.
		\item \textit{Remove cow}, which is an action for a positive test. Animal level.
	\end{itemize}
	\item \textbf{Time-Scales}\\
	A direct consequence of the events' definition, the time-scales refer again to a single hierarchical level of the system.
	\begin{itemize}
		\item BVD transiently infectious period (recovery). Animal level.
		\item Maternal antibody protection period. Animal level.
		\item Pregnancy period. Animal level.
		\item Animal movement event's period. Farm level.
   \end{itemize}
	\item \textbf{Farm Types}
		\begin{itemize}
			\item Simple One Herd Farm		
			\item Small One Herd Farm			
			\item Slaughterhouse
			\item Well
		\end{itemize}
	\item \textbf{Network Entities}
		\begin{itemize}
			\item Farm manager
			\item Market
		\end{itemize}
\end{enumerate}

\section{Process Overview and Scheduling}

\begin{figure*}[h!]
		\centering
		\includegraphics[width=\linewidth]{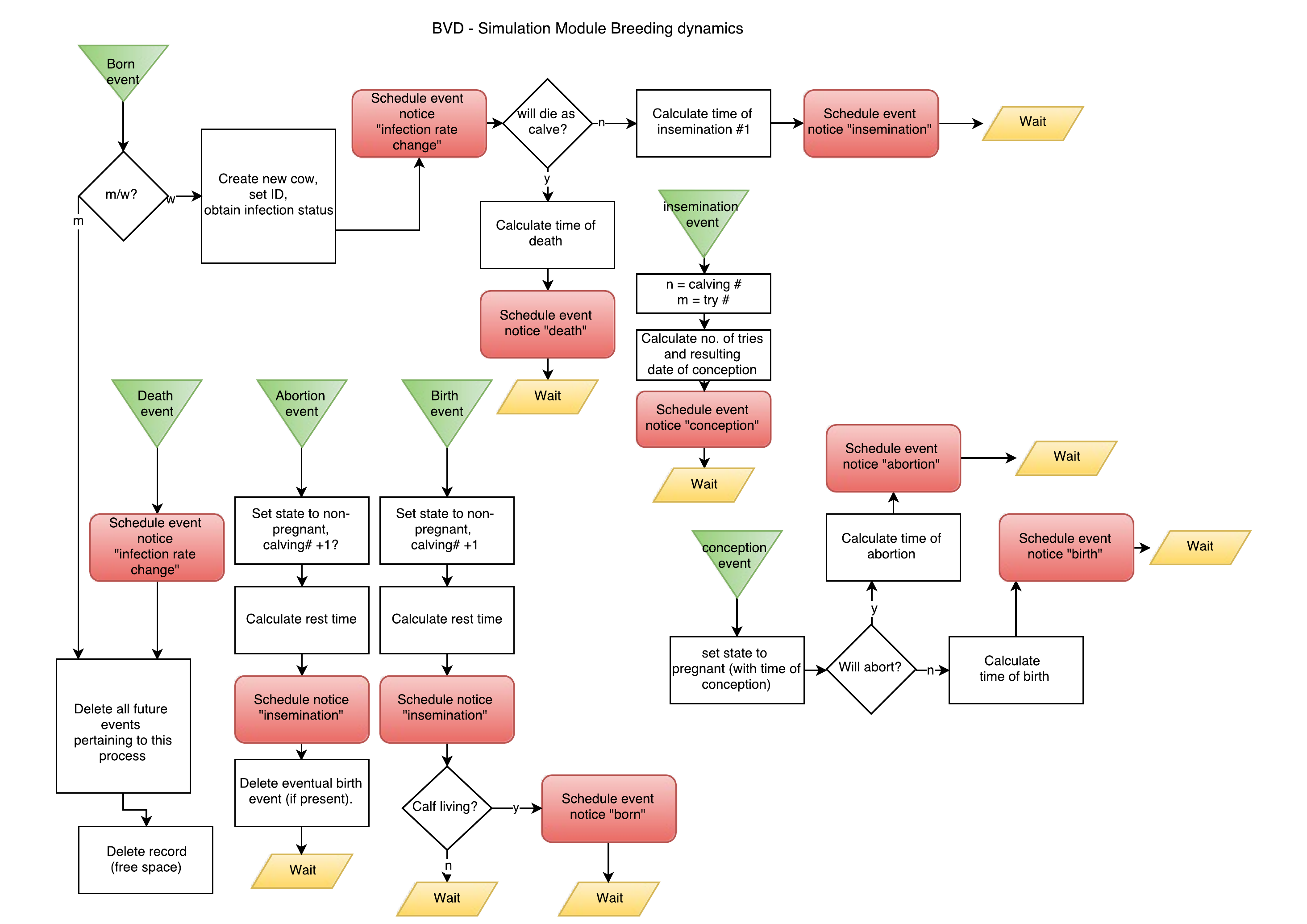}
		\caption{The simulation's breeding module in a reduced, flow chart version. All instructions refer to the animal level.}
		\label{fig:BVD_breed_mod}
	\end{figure*}

\begin{figure*}[h!]
		\centering
		\includegraphics[width=\hsize]{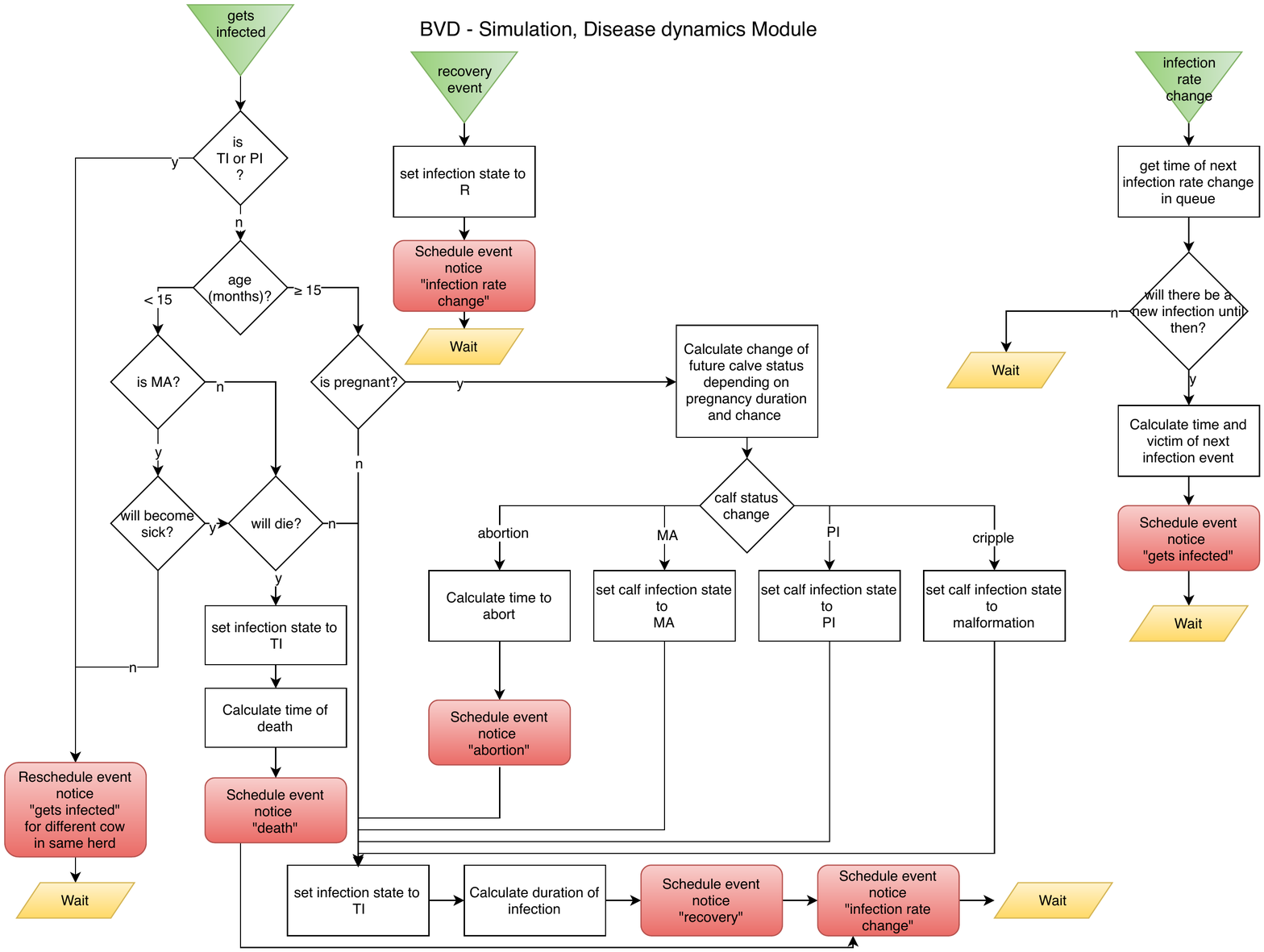}
		\caption{The simulation's infectious module in a reduced, flow chart version. All instructions refer to the animal level.}
		\label{fig:BVD_inf_mod}
	\end{figure*}

\begin{figure*}[h!]
	\centering
	\includegraphics[width=0.9\linewidth]{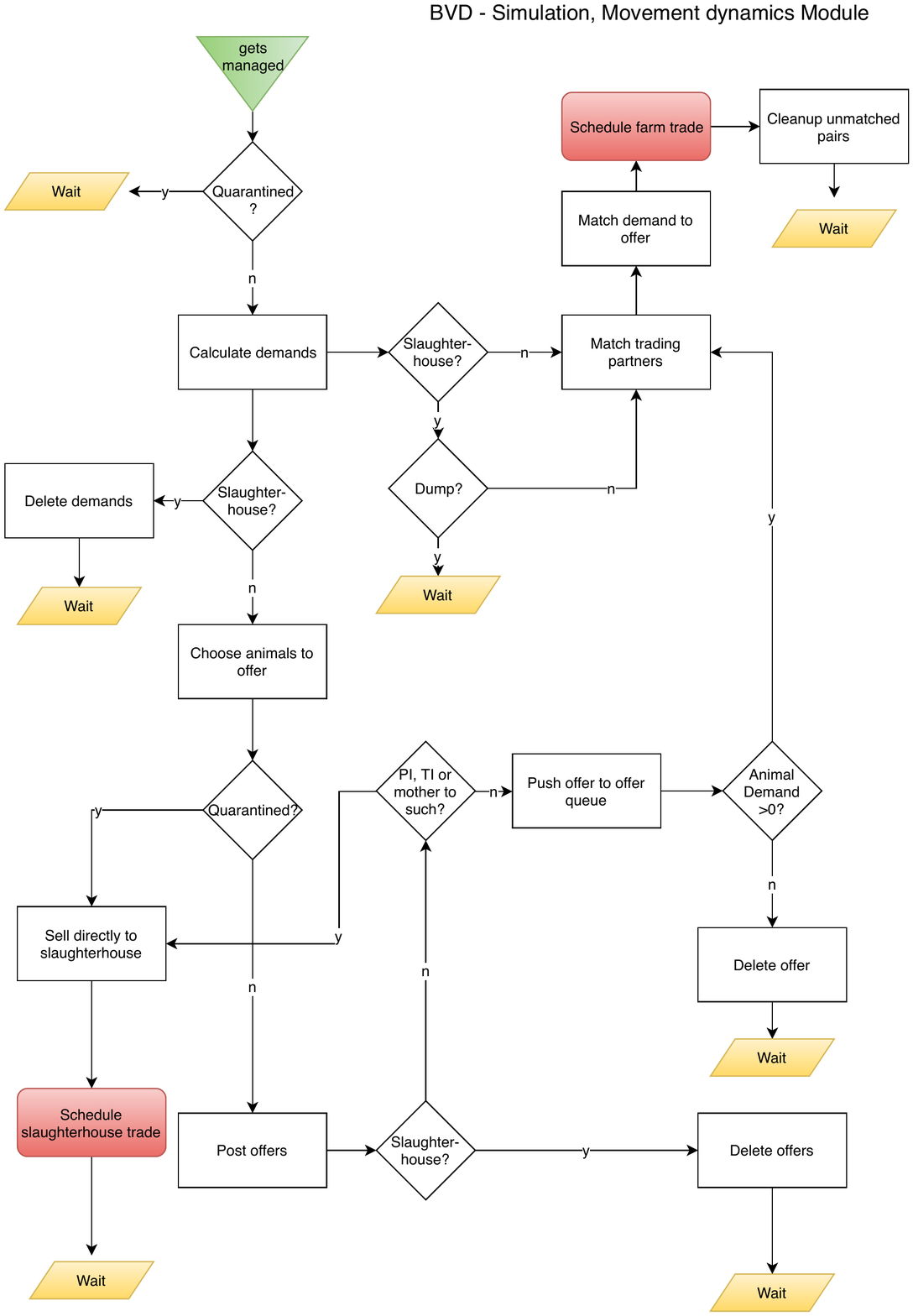}
	\caption{The simulation's movements' module in a reduced, flow chart version. All instructions refer to the farm level.}
	\label{fig:BVD_movs_mod}
\end{figure*}

\begin{figure}
	\centering
	\includegraphics[width=0.7\linewidth]{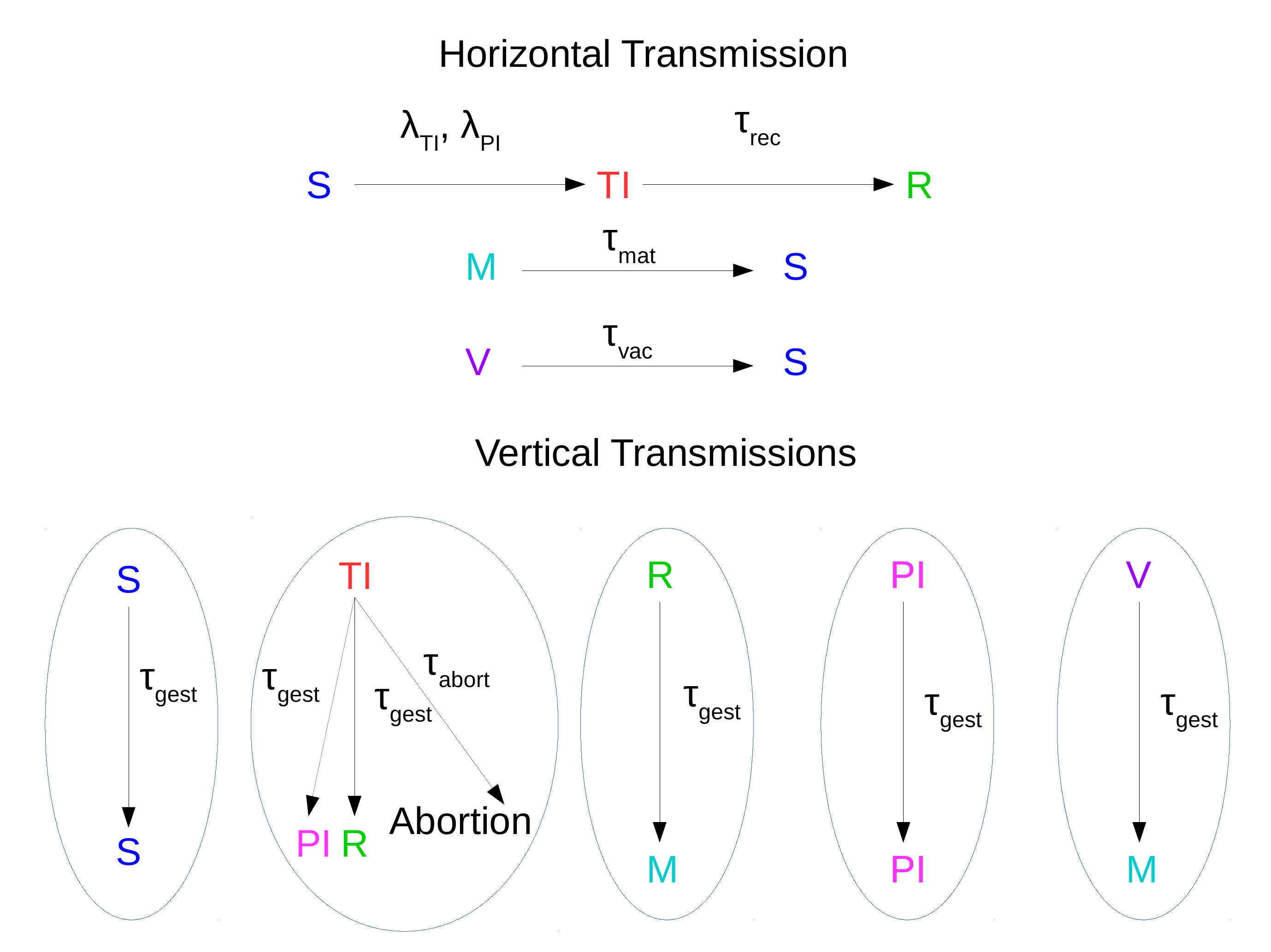}
	\caption{Horizontal and vertical transmissions of BVD. $\tau$\textsubscript{gest}, $\tau$\textsubscript{abort}, $\tau$\textsubscript{rec}, $\tau$\textsubscript{mat}, and $\tau$\textsubscript{vac} denote the gestation period, the time elapsed between the infection and an abortion, the average recovery time, the maternal antibody effect period, and the duration of protection by vaccination, respectively. $\lambda$\textsubscript{TI} and $\lambda$\textsubscript{PI} correspond to the transmission rates of S to TI from a contact between a susceptible animal and a TI or a PI animal, respectively. M refers to calves with temporary immunity being born from immune cows or heifers prior to their conception and V denotes vaccinated animals.}
	\label{fig:BVD_horiz_vertic}
\end{figure}

As an overview of the code's processes and scheduling flows we distinguish three intertwined modules as seen in Figs.~\ref{fig:BVD_breed_mod}, \ref{fig:BVD_inf_mod}, and \ref{fig:BVD_movs_mod}: one for the breeding mechanism, one for the infectious mechanism and one for the management protocol, respectively. We further present a short set of serial instructions, which illustrates the flow of the model as a whole according to a \emph{priority queue} containing all the scheduled events. We finally illustrate in Fig.~\ref{fig:BVD_horiz_vertic} the vertical and horizontal (i.e. by birth or contact respectively) infectious transmission flow as modelled in the simulation. Everything takes place in (floating point) continuous time, in which events take place in discrete points in time.

The events in turns trigger one another in the spirit of the event-driven paradigm \cite{FIS13}:
\begin{enumerate}
	\item START
	\item Initialize system (farm and animal variables).
	\item Schedule future events for initialized animals.
	\item Log the system's (initial) state.
	\item Execute the queue's events of any of the 4 system levels (system, farm, herd or animal) until either the queue is empty or the \emph{stop} event (equivalent to the specified end time) is reached.
	\item Log every event after its execution.
	\item STOP
\end{enumerate}
To give a brief explanation of the various actions and objects in Figs.~\ref{fig:BVD_breed_mod}, \ref{fig:BVD_inf_mod}, and \ref{fig:BVD_movs_mod}, the green triangles represent the initiation of an event, the rhombuses a binary query, the rectangles operations within the scope of an event, the red ovals the scheduling of an event and the yellow parallelograms a frozen state awaiting to be initiated by an event. Further, for each of the modules and across them, we can discern distinct submodules which feed each other to form the module of the corresponding figure.

\clearpage

In Fig.~\ref{fig:BVD_breed_mod} we take the example of a birth event. Such an event will eventually lead to the scheduling of an insemination for the newborn in its adult life. Until that age is reached the event is put on hold. In turns, once the insemination is executed it leads to the scheduling of a conception event within a number of tries (stochastically determined). If some try is going to be successful the conception event is scheduled, which is again put on hold until the system time reaches its execution time counting from the time of the insemination event's execution. Similarly and serially, the execution of the conception event at the appropriate system time will lead to the scheduling of either a birth or an abortion event (within some stochastically determined gestation period from the conception event's execution time) and then be put on hold. Finally, when the time for a new birth event has been reached, the circle between two birth events will have been completed.

\clearpage

In Fig.~\ref{fig:BVD_inf_mod} we present an example for an infection event. It is initiated in a farm and, depending on the age group of an animal, after several operations and queries on the animal's state will eventually lead to the scheduling of a recovery event and to the scheduling of a change of the infection rate (farm level event). Both are then put on hold. The former, once executed will in turns lead to the scheduling of the change of the infection rate of the farm again and then be put on hold. Eventually, the animal will die by the execution of a death event, scheduled from the module of Fig.~\ref{fig:BVD_breed_mod}. Once this latter event is executed one more infection rate changing event will be scheduled. Meanwhile, depending on the availability of susceptible animals in a farm, the existence of an infected animal (TI or PI) in a farm will trigger a new infection event to be scheduled according to the value of the infection rate for the farm in question at the time when the infected animal appears in it.

\clearpage

In Fig.~\ref{fig:BVD_movs_mod} the only initiated event is the managing one. For a certain farm, the managing event will go through the demands of the farm in animals (so as to keep its population constant), calculate them and depending on the farm's status of being quarantined or not might schedule a direct trade to the slaughterhouse in the former case (the only possible trade for a farm in case it has been quarantined). Then it is put on hold until the predefined time of its execution. Then the farm will be managed in its next period according to its determined value from the system parameters.

Figure~\ref{fig:BVD_horiz_vertic} summarizes the horizontal and vertical mechanisms of transmission. Horizontally, we consider an infectious transmissions of susceptible animals within a farm, owing to their (random) contact with either of the infectious animals TI or PI with corresponding infectious transmission rates $\lambda$\textsubscript{TI} and $\lambda$\textsubscript{PI}. Eventually, the infection will spontaneously lead to a recovery within a recovery period $\tau$\textsubscript{rec}. Furthermore, animals with a temporary immunity acquired from their parent (maternal antibodies) lose this immunity spontaneously within a period $\tau$\textsubscript{mat}. In addition, animals that have been vaccinated retain their immunity for a period $\tau$\textsubscript{vac}. Vertically (i.e. through breeding), a susceptible cow or heifer throughout its gestation period $\tau$\textsubscript{gest} will give birth to a susceptible animal. A cow or heifer, which has recovered prior to its conception, will give birth to an animal with temporary (maternal antibody) immunity after its gestation period $\tau$\textsubscript{gest} has elapsed. A PI cow or heifer will always give birth to a PI after its gestation period $\tau$\textsubscript{gest}. Finally, a cow or heifer that will become temporarily infected during its early pregnancy stages will give birth to a PI animal. Conversely, if it will get infected in its later pregnancy stages it will give birth to a recovered animal after time $\tau$\textsubscript{gest}. Moreover, there is a chance that the embryo will be aborted within some stochastically determined abortion period $\tau$\textsubscript{abort}.

\section{Design Concepts}

In this section we summarize the concepts permeating the design of the simulation.

\subsection{Basic Principles}

After initialization the simulation flow is executed according to a \emph{priority queue} of the simulation's scheduled events. This is a LIFO (`last in, first out') container (i.e. a data structure with specific access rules) essentially representing a queue with its elements being sorted pairwise first according to some criteria and then according to its LIFO principle \cite{SKI98,THU17}. The priority criteria were

\begin{enumerate}
	\item Causality priority. This means that given two events scheduled for execution at times $t_1$ and $t_2$ with $t_1 < t_2$, the event corresponding to $t_1$ would be sorted to be executed before that corresponding to $t_2$, even if the one of $t_2$ was stored in the queue structure before that of $t_1$.
	
	\item Event priority. This sorting follows the list sorting presented for the events in section \ref{sec:St_vars}, which has a basis relevant to the BVD disease biology and to the agricultural system's structure. Events with a higher priority are sorted to be executed before ones with a lower priority.
\end{enumerate}

Due to the sequential triggering of events in an event-driven simulation a priority queue container is the natural data structure to employ for such computations \cite{FIS13,SKI98}. Such a structure has also been implemented in a patch-based scheme \cite{ALL08,BRA12} simulation for evaluating BVD within the Irish agricultural system \cite{THU17}.

\subsection{Stochasticity}

All events pertaining to breeding, infections, recoveries, testing, vaccinations and animals selected to be classified in a certain group are executed at times and have outcomes drawn from either a uniform or a triangular distribution (float or integer depending on the application). Naturally, the stochastic times are drawn in a way which respects causality. For the choice of either one of the two aforementioned random number distributions made, the expert opinion of \cite{GET15a,GET18} was utilised. All the random generator calls employ the Mersenne-Twister algorithm as implemented in the GNU Scientific Library \citep{GNU18}. For the infectious waiting times an exponential random distribution was employed due to the \emph{Markovian} nature of the infection events.

\subsection{Sensing}

Overall, the dynamics of the animal movement network are governed by three intertwined factors: the farm manager, the market and the independent mechanisms (testing and end of life cycle) that dispatch an animal to the slaughterhouse.

Each farm of the network has a farm manager which posts its demands and offers in animals to a central entity called `the market`, which in turns decides how the trading partners (farms) are going to be distributed based on the posted demands and offers. The offer and demand of cattle therefore implicitly dictates largely the connectivity evolution of the movements' network in time. Furthermore, animals that are identified as PI or that end their life cycle and are scheduled for removal also dictate a farm's connection to the slaughterhouse.

Given the aforementioned ways with which the agents in the network evolve, their movements can provide information about an epidemic both through `infection tracing' (i.e. tracing an infection of nodes back to a source node) or through `contact tracing' (i.e. identifying all the possible contacts of nodes for control strategies) \cite{KEE05}.


\section{Initialization}\label{sec:init}

The system is initialized with all links being inactive, i.e. at the snapshot corresponding to $t=0$ the cattle movements' network has no edges.

\begin{enumerate}
	\item \textit{Setting up the farm size distribution}
	
	A farm list is read from an input csv file with two columns of integers and the total number of animals and farms (including wells and slaughterhouses
	separately from the rest of the farms) is counted at this stage. The farm distribution is created.
	
	\item \textit{Setting up the farm infectious levels, the animal count and the respective age distribution}
	
	From the total number of farms the amount of those which are going to be PI-infected and PI-free is determined
	stochastically with a threshold provided by the ini file. Subsequently the animal count and the age distribution (triangular)
	for every animal of the farm are determined.
	
	\item \textit{Setting up the types of farms}
	
	All the types of farms are initialized iterating through the farm size distribution. This effectively determines which of 
	the non well or slaughterhouse farms are going 
	to be of the small or simple one-herd type. The former does not include annual replacement on the herd, while the latter does. 
	The threshold of animals which determines the type of the farm is defined from an ini file.
	
	\item \textit{Farm initialization}
	
	For every accessed farm its corresponding number of animals is initialized.
	
	\item \textit{Animal initialization}
	
	For every initialized animal multiple parameters including its age, sex, health status (S, TI, PI, R) and future events are determined. The future events are discerned into those for male and those for female animals separately.
	
	\begin{itemize}
		\item For male animals their transfer to the slaughterhouse is scheduled at a time equal to their life expectancy. This life
		expectancy is determined by assigning different probability margins for three age cohorts of the animals. The greatest weight
		lies in the second and third, which also contain larger maximal life expectancies than the first cohort (up to around 2 years).
		More details at the birth event.
		
		\item For female animals their first insemination age is determined and the time for calving as well. If they are not of the right
		age to calve (i.e. produce offspring), then their insemination is scheduled within a time from when they will come of age to calve plus a uniform number between 0 and 
		the minimum pregnancy duration. If they are of the right age to calve, their labour is scheduled within a time ranging from the current
		time and the minimum pregnancy duration. Subsequently to the right calving age, the health status of the calves to be born is determined
		based on the health status of their mothers.
		\begin{enumerate}
			\item If the mother at this stage is TI then the calf can only be PI or an an abortion will occur. This is a stochastic outcome
			of the parameter settings for the infection during the first period of pregnancy. See tables \ref{tab:embryo_outcome_from_TI} and \ref{tab:gen_abort_stage}
			\item If the mother at this stage is PI then the calf will be a PI.
			\item If the mother is S or R at this stage the calf is set to be S (potential MA protection is provided upon the call of the birth event 
			if the mother is R).
		\end{enumerate}
	\end{itemize}
	
\end{enumerate}

\section{Submodels}\label{sec:Submodels}
In this section the functionality of the different modules of the simulation is described in detail. In spite of the fact that effort was made to set clear borderlines among the different modules there is occasionally considerable overlap as will be apparent in what is to come. The specific choice of distributions and parameters used follow \cite{GET15a}.

\subsection{Breeding Cycle}\label{sec:breed}
Naturally, this mechanism concerns only female animals (cows and heifers). It is assumed that cows and or heifers will always give birth to a single animal.
\begin{enumerate}
	\item \textit{Insemination}
	
	Specified in time upon initialization for non-calf females, the insemination is the event which can trigger a conception, and in turns a birth, and a vaccination for an animal already in the system.
	
	Apart from the initialized cattle, the first insemination is scheduled upon the birth of surviving female calves between one and a half and two years (see Tab.~\ref{tab:triang_distrs}).
	
	For cows and or heifers remaining in the system and taking part in the calving process (i.e. regardless of whether they gave a birth or had an abortion, their carriage was counted as a calving) the next insemination is scheduled during their resting time after they gave birth or had an abortion. This resting time is determined by drawing random numbers from a triangular distribution with an upper limit of roughly a month and a lower of four months (see Tab.~\ref{tab:triang_distrs}).
	
	Regarding the triggering of a conception, an insemination will determine whether and when a conception would be scheduled depending on the animal's age. In any case each animal gets up to four chances to get inseminated, each with increasing probability of success. The waiting times between successive inseminations are determined by drawing randomly from a triangular distribution with an upper limit of 18 days and a lower of 24. The setup of the model is such that heifers will always be impregnated within the four insemination attempts, while a $0.08 \%$ chance of infertility exists for the rest of the cows. In this last case the cow is declared to be infertile after the four attempts' time has passed plus a uniformly drawn random number between 0 and 14 days, and then the registering of it for sale to the slaughterhouse takes place instead of a conception event within a day from that time.
	
	Regarding the case of an active vaccination strategy, provided that the effect of a potential previous vaccination has worn out, in the case that this effect's expiry took place before a minimum time which must elapse between the insemination and the vaccination (set to a month by default), then a vaccination event is scheduled by the insemination upon the insemination's scheduling.
	
	\item \textit{Conception}
	
	This event can only be triggered from an insemination and its outcome is either the scheduling of a birth or the scheduling of an abortion. In any case the survival and state of the embryo are determined
	by the state of the mother at this stage (tables \ref{tab:embryo_outcome_from_TI} and \ref{tab:gen_abort_stage}). In particular:
	
	\begin{itemize}
		\item If the mother is transiently infected at this stage the embryo will either become a persistently infected animal or it will even more likely be aborted, in which case its abortion is scheduled within the next two days.
		
		\item If the mother is persistently infected at this stage the embryo will definitely reach birth and will be a persistently infected animal as well.
		
		\item If the mother is susceptible the embryo will be susceptible if a birth is scheduled for it.
		
		\item If the mother has any sort of immunity (i.e. temporary from a vaccination or permanent from an infection), the embryo will acquire maternal antibody protection upon its birth, if the birth is realized.
	\end{itemize}

	At this point the realization of a birth or an abortion (if the latter has not already been scheduled due to the mother being transiently infected) is determined by drawing randomly from uniform distributions within different time windows of the gestation period. These windows were defined as follows with a probability up to $12\%$ and are biologically motivated:
	
	\begin{itemize}
		\item The first two months (abortion).
		\item The second to the third month (abortion).
		\item The third to the fourth month (abortion).
		\item The remaining time up to the 210\textsuperscript{th} day of the pregnancy (abortion).
		\item The time from the 280\textsuperscript{th} to the 292\textsuperscript{nd} day of the pregnancy (birth).
	\end{itemize}
	
	For the exact parametrization of the abortion probabilities at a particular pregnancy stage see Tab.~\ref{tab:gen_abort_stage}.
	
	\item \textit{Birth}
	
	This event is either triggered for the initialized animal at the beginning of the simulation or through a conception event, as long as an animal has not been declared infertile 
	as previously explained.
	
	The history of births a cattle has had is the indicator of how many times it will continue being inseminated and therefore possibly reach a birth event, a feature which was already anticipated at the insemination's
	description. In the model this history is defined through a calving number at the creation of an animal. This number is determined by drawing randomly from a triangular distribution with an upper limit of 3 
	and a lower of 5. Therefore, the maximum number of calvings a cow can have is 5 and in every birth event a unit is subtracted from its calving number (Tab.~\ref{tab:triang_distrs}). Once its calving number has been spent, i.e. becomes zero, the cow is sent to the slaughterhouse within one day of its last labor instead of having its rest time scheduled after calving (see the insemination cycle regarding the rest times). At this stage the cow's labor is recorded in its birth history.
	
	Having dealt with the birth history of the cow in labor, the first thing that is specified upon birth is whether the embryo is a stillbirth. All further actions in respect to a birth event assume that the 
	birth is not a stillbirth.
	
	At first, the health statuses of the newborn calf and the mother are determined, the birth of the calf is noted in time and the mother is declared to be non-pregnant. The possible outcomes read thus:
	
	\begin{itemize}
		\item If the mother is susceptible then the newborn will also be susceptible.
		\item If the mother has any sort of immunity (i.e. permanent or temporary) acquired prior to the conception, then the newborn will acquire a temporary immunity inspired biologically by maternal antibody protection. The duration of this immunity is determined in days by drawing random numbers from a triangular distribution with an upper limit of 180 and a lower of 270, corresponding roughly to 6 to 9 months (Tab.~\ref{tab:triang_distrs}).
		\item If the mother has acquired permanent immunity during the pregnancy and this stage (of birth) has been reached, then this means that the calf will either be permanently infected (early pregnancy stages) or will have acquired permanent immunity (later pregnancy stages). See Tab.~\ref{tab:embryo_outcome_from_TI}.
		\item If the mother gives birth to a malformed calf, regardless of her status the model assumes that this is equivalent to the death of the calf (it is being directly put down).
	\end{itemize}
	
	Next and depending on the specified health status of the calf, its survival or not as a calf is determined.
	
	\begin{itemize}
		\item If it is a persistently infected calf then its theoretical absolute lifetime maximum is set to 10 years according to field observations. For approximately the three quarters of the observed cases,
		four cohorts are defined for the lifetime of such calves corresponding to their four first years of life. A fifth cohort accounts for the rest of the cases. Finally, the lifetime of the persistently infected
		calf is determined by drawing randomly from uniform distributions respective to each cohort, after the animal has been assigned to a particular cohort by drawing randomly from a uniform percentage distribution in the following manner (see also Tab.~\ref{tab:PI_death_age}):
		\begin{enumerate}
			\item A lifetime of up to a year with a $50\%$ probability.
			\item A lifetime of a year up to two years with a $17 \%$ probability.
			\item A lifetime of two years up to three years with a $5 \%$ probability.
			\item A lifetime of three years up to four years with a $1.5 \%$ probability.
			\item A lifetime of four years up to ten years with a $26.5 \%$ probability.
		\end{enumerate}
		
		\item In all the rest of the cases the calves are distributed in three mortality age cohorts of their category by drawing randomly from a uniform percentage distribution. Each cohort has a corresponding lifetime probability, being drawn randomly from uniform distributions once more, defined in the following:
		\begin{enumerate}
			\item $[0, 2.5)$ days with a $2.5 \%$ probability.
			\item $[2.5, 182.5)$ days with a $10 \%$ probability.
			\item $[182.5, 365)$ days with a $1.5 \%$ probability.
		\end{enumerate}
		
	\end{itemize}
	
	Afterwards the sex of the newborn is determined (see Tab.~\ref{tab:unif_distrs}). If the newborn is female and it has been determined to survive its first insemination age, then its first insemination age is scheduled by drawing randomly from a triangular distribution with an upper limit of 480 days and a lower of 600. 
	
	If the newborn is a male, additionally to its survival conditions set before, its life expectancy is scheduled as outlined for the male animals at initialization. The male newborns are distributed in the cohorts by drawing randomly from a uniform percentage distribution. These three cohorts have a corresponding lifetime probability, each of which is defined by drawing randomly from a triangular distribution. At the end of its lifetime each male animal is sent to
	the slaughterhouse. The probabilities to be distributed in each cohort and the
	two limits of the corresponding triangular distributions (\ref{tab:triang_distrs}) are as follows:
	
	\begin{enumerate}
		\item Upper limit $0$ and lower $30$ days with a $30 \%$ probability.
		\item Upper limit $170$ and lower $250$ days with a $35 \%$ probability.
		\item Upper limit $450$ and lower $750$ days with a $35 \%$ probability.				
	\end{enumerate}
	
	If the newborn is a female and it survives at least until its first insemination age, then an insemination is scheduled for it at that time.
	
	Lastly, in case an ear tag strategy is in effect the testing time of the newborn is scheduled for a time being drawn randomly from a triangular distribution with an upper limit of 4 days and a lower of 30 (first month of life).

	\item \textit{Abortion}
	
	This event essentially checks that the animal in question is indeed female, cancels a possible scheduled future birth or abortion (see explanation in the infection scheme), counts
	the abortion as a calving if the time elapsed since the last conception event is larger than 240 days and schedules the resting time of the animal until its next insemination (if applicable 
	--see insemination's description).
	
	\item \textit{Death}
	
	Any of the \textit{death, culling} or \textit{slaughter} events encountered are treated similarly and signify the removal of the animal in question (and its associated scheduled events) from the system.
	
	\item \textit{Rest time after abortion}
	
	This event has a dual role. Apart from scheduling the rest time of the animal in question as described in the insemination, it also sets the criterion which signifies the end of the breeding cycle of the animal.
	This criterion is that the calving number of the animal has reached to zero. In this case its selling to the slaughterhouse is scheduled within half a day.
	
\end{enumerate}

\subsection{Infection Cycle}

\begin{enumerate}
	\item \textit{Infection}
	
	An infection happens at the animal level for susceptible animals and is dependent on the number of its infected neighbours, i.e. the number of TI and PI animals. 
	The neighbourhood of an animal is defined to be any member of the herd with an equal contact probability and infections in a specific time $t$ take place with a certain instantaneous, stochastic infection rate of the form found in \cite{VIE04}:
    \begin{equation}\label{eq:infect_rate}
	    \lambda _{l}(t) = \beta\text{\textsubscript{PI}} \frac{\text{PI}_{l}(t)}{N_{l}(t)} + \beta_{\text{TI}}\frac{TI_{l}(t)}{N_{l}(t)} + \sum_{m \neq l} \beta_{\text{m,l}}\frac{\text{PI}_{m}(t)}{N_{l}(t)N_{m}(t)},
	\end{equation}
	where $\beta$\textsubscript{PI} and $\beta$\textsubscript{TI} are the BVD transmission coefficients with inverse time units, PI\textsubscript{l}(t), TI\textsubscript{l}(t) and N\textsubscript{l}(t) the PI, TI and total animal number in the herd $l$ at time $t$, $\beta$\textsubscript{l,m} the BVD transmission rate from the PI animals of any other herd $m$ to $l$ and PI\textsubscript{m}(t) and $N_{a,t}$ the number of PI animals in herd $m$. While the $\beta$ coefficients are fixed throughout the simulation, the TI, PI and total herd numbers are generally stochastically changing for every time $t$, thus the characterisation of the rate $\lambda _{l}(t)$ as stochastic. If the rate $\lambda _{l}(t)$ is multiplied with the population of S of the herd (number of infection candidates) for time $t$ then the resulting rate represents the \textit{total infections per time unit} $a(t)=\lambda _{l}(t) \times S(t)$.
	
	It is clear from the summation term in equation \eqref{eq:infect_rate} that infections can also occur across herds as well (as only one herd per farm is simulated in the scope of this study, this is of no concern), but only due to the infectiousness of the PI animals as in \cite{VIE04}. This introduces subnetwork dynamics within the node, where each herd is a node and can influence its neighbouring (i.e. within the farm node) herd nodes' infectious states. The TI contribution is negligent in this case of infection in-between herds.
	
	Regarding the infectious process, we define it to be a random, Markovian event within the herd $l$. Due to the fact that the average waiting time for such events from one arbitrary time point $t_1$ to the next $t_2$ is exponentially distributed \citep{KAM03}, the total infections per time unit $a$ are constant in that time window (and thus its factors $S$ and $\lambda _{l}$) and serves as the rate parameter of the exponential distribution from which the probability to observe a waiting time $\tau_w$ in the said time frame is drawn. Said differently, the average waiting time between two successive infection events at $t_1$ and $t_2$ ($t_1 < t_2$) for a constant and positive, non-zero infection rate coefficient $a$ in this time interval is
	\begin{equation}\label{eq:wait_time}
	\langle \tau_w \rangle = a \int_{0}^{\infty} e^{- at}t \,dt = \frac{1}{a}
	\end{equation}
	with the integral's limits corresponding to the domain on which time is defined.
	
	We stress that the average waiting time from one infection to another is defined for constant total infections per time unit $a$ in the time window $[t_1, t_2]$. This $a$ is in turns calculated after the susceptible population in the herd in question and equation \eqref{eq:infect_rate} for every time an infection rate changing event takes place, i.e. for any event of priority equal or lower to that of \emph{birth} as ordered in the event list. Thus the average waiting time $\langle \tau_w \rangle$ changes with every change of the instantaneous infection rate $\lambda _{l}(t)$. By examining the two marginal cases of the total infections per time unit $a$ between two successive infection events at $t_1$ and $t_2$ the necessity of the definition of $a$ becomes evident.
	
	\begin{itemize}
		\item For $a \gg 1$ or $t_1 \rightarrow t_2$ then $\tau \rightarrow 0$, which means that the waiting time is negligible if the distinct infection events are very close to each other, if the pool of susceptible candidate victims is enormous in that time window ($S \gg 1$) or conversely if the instantaneous infection rate of \eqref{eq:infect_rate} assumes a high value in $[t_1,t_2]$, $\lambda _{l} \gg 1$.
		\item For $a \rightarrow 0$ then $\tau \rightarrow \infty$. This means that if the pool of the susceptible population is zero and the instantaneous infection rate is not zero (e.g. if the S portion of the population has been depleted and only R, TI and PI remain), then the waiting time between two successive infection events will be infinite as (for all else remaining constant) there will be no available candidates to infect.
	\end{itemize}
	
	It follows that, although the infection event
	refers to the animal level, the \emph{infection rate changing} event refers to the herd level and therefore its change will correspond to a herd, not an animal. Note that if the infection
	rate is scheduled to change before a potential scheduled infection, the infection will not be scheduled as the conditions for its realization change exactly due to 
	the rate's variation. Consequently, an infection event's \emph{transition probability} is only dependent on the state variables corresponding to the time step directly 
	prior to its execution, therefore qualifying it to be termed a \emph{Markov process} \cite{KAM03}.
	
	As already mentioned, for the particular case of this model, the inter-herd transmission coefficients do not play a role as they refer to herds within the same unit (e.g. farm) and a single farm has been modeled to contain only one herd.
	
	Once an animal from a certain herd has been chosen for an infection its age plays a role in the decision on whether it should recover or die from the disease. Furthermore, if 
	the animal is a carrying cow or heifer, a stochastic decision is made on the effect of the disease on its pregnancy. The exact aforementioned cases read as follows:
	
	\begin{enumerate}
		\item \textit{Calf Animal}: A decision is made on its survival from BVD (see Tab.~\ref{tab:calf_TI_results}):
		
		\begin{itemize}
			\item If it survives, its recovery is scheduled after its infection duration.
			\item If it does not survive, its death is scheduled after its infection duration.
		\end{itemize}
		
		\item \textit{Non-Calf Animal}: Its sex is determined. If the animal is a cow or a heifer the outcomes of BVD on the embryo are the following depending on the
		stage of the pregnancy:
		
		\begin{itemize}
			\item It will be persistently infected (early stages).
			\item It will be aborted (early to mid stages).
			\item It will be malformed, which in this model is killed immediately (mid to later stages).
			\item It will have lifelong immunity (mid and mostly later stages).
		\end{itemize}
		
		\noindent
		Note that the pregnancy stages were only qualitatively described here and correspond to four successive cohorts in time. Each cohort has its own probability definition
		for the possible embryo outcome. For details see Tab.~\ref{tab:embryo_outcome_from_TI}. Furthermore, an important
		assumption made for carrying cows and heifers during the infection is that should an infection-caused abortion be defined to occur sooner than an already scheduled abortion originating from 
		a conception event, then the conception-caused abortion will be invalidated and the infection-caused abortion will take its place in the event schedule. Moreover, three marginal 
		cases of the abortion in respect to the birth scheduling are distinguished so as to avoid conflicts. Specifically:
		
		\begin{itemize}
			\item If an infection-induced abortion is to take place after a scheduled birth, then the abortion is executed immediately.
			\item If an infection-induced abortion is to take place before a scheduled birth, then the abortion is scheduled as planned with all the considerations so far outlined.
			\item If an infection-induced abortion is to take place simultaneously with a birth, then the birth prevails.
		\end{itemize}
		
		Regardless, the non-calf animal is not modeled to die from the infection, thus its recovery is scheduled after the duration of the infection.
		
	\end{enumerate}

	\item \textit{Infection rate change}
	
	Any of the events \textit{birth, death, end of MA, infection, recovery, trade, remove cow, slaughter, culling, vaccinate} or \textit{end of vaccination} change either the herd population, the TI or the PI population, or change the neighbors of any animal within the herd. This means that these events would change the infection rate \eqref{eq:infect_rate} directly or its effect on the population of susceptible animals.
	
	The trade and the rest of the infection rate changing events are handled separately. This is because in the former case the scheduled events are transferred along with the animal to the destination herd/farm, while
	in the latter all the scheduled events for the animal pertain to its herd (also the farm in the modelling so far).
	
	\item \textit{Recovery}
	
	This event simply depopulates the TI group and transfers the output to the R group at the scheduled times of recovery for the corresponding animal.
	
	\item \textit{End of maternal antibody protection}
	
	This event signifies the end of the maternal antibody protection from BVD for calves drawing random numbers from a triangular distribution ranging up to three quarters of a year (see Tab.~\ref{tab:triang_distrs}).
	
	\item \textit{Vaccination scheduling, duration and expiry}
	
	If a vaccination strategy is in effect, the first vaccination of a surviving calf is determined upon its birth for its 186\textsuperscript{th} day of age. The effect of this vaccination for a susceptible animal, (i.e. a transition in the R group with a note that the immunity is temporary), provided it is
	successful, is set for a year (365 days) from the ini file, upon which time a compensation between the R and S groups will take place for the particular animal. If the animal is not susceptible, the vaccination will simply have no effect on the health status of the animal, but its next vaccination will be scheduled in one year from the current time. The only reason for a vaccination not to take place in precisely these scheduled time frames, provided the animal in question is alive, is that the vaccination scheduling overlaps with the minimum distance from the insemination (see \ref{sec:breed}).
	
\end{enumerate}

\subsection{Testing Schemes}

We consider both testing an animal for BVD through an antigen test (the so called \textit{virus test}) or by an antibody test are in effect, depending on the implemented testing strategy. The following distinctions of tests are in terms of periodicity and scope, i.e. either once in their lifetime for all animals or periodically for a sample of animals from each farm.

\begin{itemize}
	\item \textit{Ear Tag}
	
	If the \textit{ear tag} strategy is in effect, then a test is scheduled for the animal in question depending on its age and scheduled only by the birth scheme. This test concerns the antigen test (virus test) by a tissue sample (ear tag), is non periodical in the 
	lifetime of the animal and can lead to a second round of antigen test testing (blood testing).
	
	Before anything else, what is determined is whether the test is positive. This can naturally be either a true or a false positive depending on the set sensitivity threshold value from the ini file (i.e. the probability to detect a truly sick animal) and the health status of the animal. The specificity (i.e. the probability to successfully identify a non-infected animal) has been assumed to be unity (certainty) for the aims of the simulation \cite{GET18}.

	\begin{enumerate}
		\item If the test is positive, then firstly the default values are being read on whether the farm will be quarantined and for how long. This depends on the strategy implemented, but for the baseline scenario (no strategy) no quarantine is enforced. Next the animal is being registered to have been tested once with a 92\% probability and with the remaining 8\% for a second test. If the animal is to be tested only once, then it is registered for removal within a time drawn randomly from a uniform distribution between 3 and 34 days. In the case of a second scheduled test, the next test is scheduled by drawing again randomly from a uniform distribution between the same day and a maximum time for retesting the animal set from the ini file. By default this is set to two months (60 days). The second round of test is a different event than the first (`virus test' instead of `test') accounting for a blood test instead of a tissue one. If the animal is specified to be positive again, this time its removal is scheduled by drawing randomly from a uniform distribution between 3 and 23 days. Lastly, if the animal is a cow and has had offspring, all of them are declared to have had a positive mother. This is significant to determine whether the offspring of a cow corresponding to a farm should automatically also be sent to the slaughterhouse without further testing if their mother's test results is positive.
		
		\item If the test is negative the animal is simply declared to be such and in case it is not one of the initialized cows it is declared negative in the event that its mother had been positive.
	\end{enumerate}
	
	\item \textit{Young Calf Window}
	
	If the \textit{young calf window} (YCW) strategy is in effect, then a sample of animals is scheduled to be selected for testing for every farm of the system periodically. The period is defined in the ini file and is set by default to be 186 days (approximately half a year). The methodology of this approach follows \cite{CON15} and asserts that given a herd of a certain size only a random sample of its animals older than six months and younger than two years of age suffices to 
	determine the existence of infectious animals in it up to a certain limit with a confidence of 95\% through testing. That is because animals over six months of age will have, on average, lost any potential maternal antibody immunity that they may have acquired upon birth by the time of the test (especially in an agricultural system where the calves are being massively administered colostrum with their first meal) and if they are younger than two years of age they will not have, on average, started to produce offspring. Therefore, any positive tests from such a specimen would suggest that there has likely been some source of infection in the farm. In the model's case the upper limit of 20\% of sick animals in the herd (of size $N$) for a given random sample size $n$ of negatively tested animals was set, as displayed in Tab.~\ref{tab:YCW_table}. The table (the simulation's sample size is within the population limits of Tab.~\ref{tab:YCW_table}) was derived with reasoning starting from the hypergeomteric distribution (formula \eqref{eq:hypergeom}). That means that given a finite population of animals $N$, from which $n$ are randomly drawn (tested) \emph{without replacement} and $K$ are indeed positive in the population, then $k$ from the drawn ones will be positive. Formulated symbolically, this translates to the probability mass function
	\begin{equation}\label{eq:hypergeom}
		p\left( X=k \right) = \frac{ \left( \begin{array}{c} K \\ k \end{array} \right) \left( \begin{array}{c} N-K \\ n-k \end{array} \right) }{ \left( \begin{array}{c} N \\ n \end{array} \right) },
	\end{equation}
	where with $\left( \begin{array}{c} a \\ b \end{array} \right)$ the binomial coefficient is meant \cite{KRI06}. Since what is being sought with this strategy is herd immunity, the antibody test is used on the sampled animals.
	
	\begin{table}
		\begin{center}
			\begin{tabular}{|c|c|c|c|c|c|c|} \hline 
				N & $\leq$ 10 & $\leq$ 20 & $\leq$ 40 & $\leq$ 80 & $\leq$ 160 & $>$ 180 \\ \hline
				n & 8 & 10 & 12 & 13 & 13 & 14 \\ \hline 
			\end{tabular}
			\caption{Population sizes (N) and the corresponding samples (n) needed to be tested negative so as to verify a maximum of 20\% of infected animals in the population with a confidence of 95\%.}\label{tab:YCW_table}
		\end{center}
	\end{table}
	
	
	In case even one animal is identified as positive during the YCW protocol, then all the animals of the herd (and thus of the farm in the scope of the model) are scheduled within half a day to be tested according to the virus test (blood test) of the \textit{ear tag} protocol previously described.
	
\end{itemize}

\section{Animal Movements}

This is the module that builds the network and allows it to evolve. The terms \textit{trade} and \emph{movement} shall be used indiscriminately here, regardless of whether the movement is defined in the system to be an actual trade or a removal of the animal i.e. a dispatch to a slaughterhouse.

\begin{enumerate}
	\item \textit{The Manage Action}
	
	This is the first stage a farm has to go through in order to have some contact with the rest of the system. The \textit{manage action} consists of a series of actions, 
	which assess whether a farm can have any interactions with the rest of the system in the first place, whether it needs input, if it is eligible for output and of what kind should
	the input or output (animals) in question be. To that end, each farm has its own managing protocol called the \textit{farm manager}.
	
	To start with, for the running time of the system a daily management period is set by default, but this can be changed from the ini file and for the results we have assumed a weekly management period. That means that each and every farm runs its management protocol (the farm manager) in every management period of 7 days.
	
	The management protocol is comprised of a number of steps.
	
	\begin{enumerate}
		\item Check if the farm is under quarantine for selling and for buying animals. If yes, that is the end of the current management action.
		
		\item The demand of animals is calculated. This is realized by firstly ensuring that the animals to be purchased will only be pregnant cows or heifers and 
		above a purchasing margin set at the input ini file (by default this is zero). Next, the actual demand calculation takes place, which depends on the type
		of farm taken into account. Currently this would mean either a \textit{simple one herd farm} or a \textit{small one herd farm}, the difference of the
		two lying in the inclusion of herd rejuvenation or not respectively. Regardless, a quota of pregnant cows demanded by
		the farm in question for a specific herd is set upon its creation and is equal to its herd size per herd (remember that the farms in this scope include only one herd each).
		Thus, the number of requested cattle is calculated as the difference between the fixed quota and the herd size in question at the time of demand, for the case 
		that this difference is above zero. Naturally, if the farm type is of the simple one herd, the fixed quota number is reduced by the farm's replacement percentage.
		
		\item The farm's demand is registered to the market (see further below in the same section), which decides on  which two farms (of any farm type) will be the 
		exchanging partners.
		
		\item Next the supply of animals is calculated for the farm at hand. For that a similar methodology is followed to that for the demanded animals, with the difference 
		being that the momentary herd size should be larger than the fixed quota of animals for any number of animals to be registered for selling. 
		
		The farm manager goes through ten available animal trading criteria for each herd and groups the herd's animals to each one of them according to their sex and age status. Then it attempts to sell as 
		many of the herd's animals as possible from groups with a higher priority and disproportionately many animals when compared to the rest of the groups. These criteria are summarized in their corresponding sex, age and fertility groups in Tab.~\ref{tab:Trade_criteria}. The reasoning of categorizing the animals to such trading groups is that, should the animals to be traded be prioritized, then animals with decreased
		financial benefit against others (mainly older or infertile animals) should be traded first if possible. As implied though, it is possible to set the selling sorting criteria to be evenly distributed 
		among the different groups through a setting in the ini file or by leaving the related field empty. 
		
		Regarding priority trading criteria, at this stage it is possible to define the trading criterion 
		for the animals to be following the numbering of Tab.~\ref{tab:Trade_criteria} in descending order for each farm. Animals from groups with the same priority numbering are drawn for filling the selling group with a weight proportional to the group size they belong to. This last point is also the distribution rule for the offered group for sale when there is no prioritizing of the trade). The way to activate this priority in animal selling is to fill the related field of the ini file with the value `OldCowsFirst'.
		
		\item Provided the farm is not under quarantine, similarly to the farm's demand, its animal supply is registered to the market, which decides on  which any two farms will be once again the exchanging partners. 
		
		\item If the farm is under quarantine its animal supply is met by the slaughterhouse's demand.
		
		\item Before the management cycle for the farm ends the farm's registered demands and offers (in the form of queues) which were not directed to a trade by the market are fulfilled. This is achieved in two different ways depending on the related setting from the ini file. 
		
		The first (called `dump') consists of matching all of the farms' demands from the well farms and conversely, all of the farms' offers from the slaughterhouse. It follows logically that the demands and offers in question cannot come from the slaughterhouses or the wells as they serve as drains and sources of the system respectively. Furthermore, self-trade is inhibited as well as direct trade from the wells to the slaughterhouses. 
		
		The second (called `demand') treats the slaughterhouses and the wells as the rest of the farms, with the logical restriction that the first cannot post offers to the market while the latter cannot post requests (demands). In case of unfulfilled trade offers the animals offered remain in their farms.
	\end{enumerate}

	\item \textit{The Market}
	
	If the market is to be distinguished from the management cycle, then it is simply to make clear that it performs the matching between the trading partners (i.e. different farms) and the demands and offers posted to it by the farm manager with queue containers. It is therefore an integral part of the management cycle, but a separate entity from the farm manager protocol.
	
	Effectively it uses a similar mechanism for both registering demands and offers of animals to match them in pairs. The criterion for the matching is defined by four factors:
	
	\begin{itemize}
		\item The trading partners (farm types) cannot be the same (self-trade).
		\item The trading partners cannot be the pair well-slaughterhouse (source to drain case).
		\item If the behavior of the well and the slaughterhouse is set to fulfill the offers and requests of the unfulfilled trades at the end of a management cycle (`dump' setting in the ini file) the farm registering a demand cannot be a slaughterhouse (see the last step of the \textit{manage} protocol).
		\item If the previous point is the opposite (`demand' setting in the ini file), all the remaining offers and requests are annulled.
	\end{itemize}
	
\end{enumerate}

\begin{table}
	\begin{center}
		\begin{tabular}{|l|l|} \hline 
			Group Type & Status \\ \hline
			1) Male calf  & Male of age $<$ 6 months \\
			1) Young bull & Male of age $<$ 17 months \\
			1) Old bull  & Male not calf and not young \\
			1) Infertile  & Female $>$ 15 months with $>$ 3 inseminations \\
			2) Old cow  & Female of age $>$ 4 years \\
			3) Heifer pre breeding  & Female animal of age 6-15 months \\
			4) Heifer ready for breeding  & Female of age $>$ 15 months with no offspring \\
			5) Calf & Female up to 6 months of age \\
			6) Pregnant  & Female of age $>$ 25 months with offspring \\
			6) Dairy cow  & Non-pregnant cow able to breed \\ \hline
		\end{tabular}
		\caption{The various age groups the animals of a herd can be grouped in to be sold according to some global selling strategy. The numbering corresponds to the ascending diminishing of prioritisation for the various selling groups assuming the OLD\_COWS\_FIRST selling strategy. Groups of the same number have the same level of priority. In case of an evenly distributed selling strategy the group numbering is rendered irrelevant. A month is assumed to have 30 days.}\label{tab:Trade_criteria}
	\end{center}
\end{table}

\subsection{Farm Types}\label{sec:farm_types}

In this section we describe the differences of the four different farm types. The threshold for farms to be either of the simple one herd or of the small one herd type can be set in the ini file according to the number of animals corresponding to a farm. The field is called `smallFarmSizeMax`and indicates the limit of animals that a farm should have to be specified as a small one herd farm.

\begin{itemize}
	
	\item \textit{Simple One Herd Farm}
	
	This is a farm containing only one herd of animals. It both offers and demands animals according to its needs, which are to preserve its population constant in every management period by comparing some quota set  upon initialization randomly from the given farm size distribution, reduced by a replacement percentage corresponding to a rejuvenation strategy of such farms, with the instantaneous population of the farm. Symbolically those needs are expressed in equation \eqref{eq:replacement} rounded to the closest integer value. The surplus sees to the term $ N_{quota} \times (1-replacement)$ being greater than $N_{instant.}$, while the opposite holds for the deficit.
	
	The replacement percentage for the one herd farm type is set by default to be $27.9\% \times 7/365$ \cite{GET18}, which roughly translates to a quarter of the herd being rejuvenated
	every year for a weekly management period (numerator) and can be altered in the given ini file. Rejuvenation means that a said percentage of animals is subtracted from the herd population's quota and is requested once per year through trades in pregnant heifers. Given the default values, it becomes evident from equation \eqref{eq:replacement} that this rejuvenation effect has an effect only for farms with an animal population above 100.
		\begin{equation}\label{eq:replacement}
	\text{surplus/deficit} = \left| N_{\text{quota}} \times (1-\text{replacement}) - N_{\text{instant}} \right|
	\end{equation}
	
	\item \textit{Small One Herd Farm}
	This farm type is similar to the simple one herd farm with the sole difference being that it does not include rejuvenation for its animals via trades. The rationale behind this is that small unit farmers will keep in general their domestic animal population constant and not renew it throughout the animals' lifetime.
	
	\item \textit{Slaughterhouse}
	
	This farm type has a dual function as a sink and as a demand farm which can be alternated in the ini file. On the one hand it can act as a sink for the simulation (slaughterhouse type demand set to `dump' in the ini file), i.e. if after all the demands in a management period have been met the market still has unmatched offered animals, those animals will be channeled to the slaughterhouse. On the other hand, the slaughterhouse can act as a small one herd farm, but only with demands (slaughterhouse type demand set to `demand' in the ini file), therefore contributing to the supply-demand mechanism of the market. Thus, it will always ascertain that all the offers that it can accommodate (`dumping capacity per type' setting in the ini file) will be met. The rest will remain in their original farms.
	
	\item \textit{Well}
	
	This farm type acts inversely to the slaughterhouse dually as a source and an offering farm depending on the settings of the ini file (same as for the slaughterhouse). On the one hand, if there are not enough offers of animals to keep the population constant it will provide them (slaughterhouse type demand set to `dump' in the ini file). On the other hand, it can also act as a small one herd farm which only offers animals,  making sure that no demand is left unmet at the end of a management period, provided the well has enough animals to offer (`number of cows in well' setting in the ini file). 
	
\end{itemize}

Note that either one of the functions of the well or the slaughterhouse are mutually activated as per the description. Therefore they can either simultaneously function as a sink and source only entity of the simulation or as small one herd farms which either only demand or only offer animals, fulfilling the unmatched offers and demands in the market to their given capacity respectively. Assuming the latter setting in effect (`demand') it is possible to include more than one well or slaughterhouse in the simulation.

\subsection{Parameter Selection}\label{sec:param_sel}
For all the parameters not concerning calibration expert opinion was used \cite{VIE04,GET15,GET18,EZA07,BIO16}. All the simulations ran for a simulation runtime of T\textsubscript{sim} = 20,000 steps with a resolution of 5 steps. As a result of the global parameters' selection (tables \ref{tab:sys_params} and \ref{tab:farm_params}), those time steps correspond to calendar days. The tests' specificity success probability was assumed to be 100\% for the needs of the simulation following expert opinion \cite{GET18}.

To start with, Tab.~\ref{tab:sys_params} contains all the parameters in regard to all the dynamics within the farm and the scope of the strategies presented in Tab.~\ref{tab:Strats}. The farms' population infectious states were initialized randomly from the given farm size distribution CSV file. Their infectious states were also randomly allocated between the PI and PI-free farms (see Tab.~\ref{tab:farm_infect}) with a 98\% bias towards PI-free farms, as explained in Tab.~\ref{tab:farm_params}. Furthermore, Tab.~\ref{tab:farm_params} contains all the relevant details permeating the type of farms, their number and their animal movement capabilities through the market.

In tables \ref{tab:calf_TI_results}, \ref{tab:embryo_outcome_from_TI}, \ref{tab:insem_results}, \ref{tab:gen_abort_stage} and \ref{tab:PI_death_age} all the probabilities concerning hard-coded infectious and breeding parameters as well as their interplay during the simulation and upon initialisation are presented. Moreover, in tables \ref{tab:unif_distrs} and \ref{tab:triang_distrs} the usage of uniform and triangular random distributions for number generation in the simulation's different processes (breeding and infectious in the former, and breeding, infectious and testing in the latter) is displayed. The latter implicates lack of homogeneity and detailed information about the interval over which the process in question takes place and its `mode' defines an `educated guess' bias \cite{KRI06}. The application details and necessity of the aforementioned tables become apparent as one inspects the various submodels listed and unravelled in section \ref{sec:Submodels}. Lastly, for all the statistical results the pseudo random number generator environment was setup as the \emph{GNU Scientific Library} dictates \cite{GNU18} was seeded with the number 2,333,600,960.

\begin{table*}
    \begin{center}
    \begin{scriptsize}
			\begin{tabular}{ccc} \hline
				Parameter & Value & Description \\ \hline
				$\beta_{TI}$ &  0.03 per time unit & TI infectious coefficient in equation \eqref{eq:infect_rate} \cite{VIE04} \\
				$\beta_{PI}$ & 0.5 per time unit & PI infectious coefficient in equation \eqref{eq:infect_rate} \cite{VIE04} \\ 
				$\Delta t$\textsubscript{TI-Ab.} & 2 days & Time between infection and abortion for a TI, pregnant cow \\
				$\Delta t$\textsubscript{Infert., Fin. insem.-Remov.}  &  14 days  &  Time elapsed between last insemination attempt and removal for an infertile cattle \\
				Calf age threshold  & 180 days & Beyond this age the animal is not a calf anymore \\ 
				Abortion as calving & 240 days & Beyond this pregnancy stage the abortion is counted as a calving \\
				TI calf death prob. & 2\%      & Probability for a TI calf to die from the infection \\
				1\textsuperscript{st} vacc. age & 186 days & First vaccination age for the animal \\
				Vacc. work prob.  & 98.5 \%  & Vaccination working probability \\
				Vacc. effect $\Delta t$  & 365 days & Effect duration of a successful vaccination \\ 
				$\Delta t$\textsubscript{Vacc.-Insem.}  & 42 days & Time interval required for a vaccination to take place before an insemination \cite{DAM15} \\
				Sensitiv. success prob. & 99\%  & Test's sensitivity success probability \\ 
				Test again prob.  &  2\%  & Probability for positively tested animals to be retested \\
				$\Delta t$\textsubscript{tests}  &  60 days  &  Time elapsed between two tests in the old regulation (strategy 2 in Tab.~\ref{tab:Strats}) \cite{LEG08} \\
				$\Delta t$\textsubscript{tests}  &  40 days  &  Time elapsed between two tests in the new regulation (strategy 3 in Tab.~\ref{tab:Strats}) \cite{LEG08a} \\ 
				$\Delta t$\textsubscript{quarant.}  &  40 days  &  Quarantine period for a farm after the new regulation \cite{LEG08a} \\ 
				T\textsubscript{ycw,1}  &  186 days  &  Periodicity of the YCW test (strategies 5a, 6a, 7, 9 as in Tab.~\ref{tab:Strats})  \\ 
				T\textsubscript{ycw,2}  &  356 days  &  Periodicity of the YCW test (strategies 5b, 6b as in Tab.~\ref{tab:Strats}) \\ \hline
			\end{tabular}
		\end{scriptsize}
		\caption{Global infectious, breeding, testing, quarantine and vaccination parameters set in the simulation per strategy (see Tab.~\ref{tab:Strats}) where applicable. The programme follows the outline of \cite{GET15a} unless explicitly stated otherwise in the table. All the testing and vaccination parameters can be controlled from the input ini file. The rest are hard-coded.}\label{tab:sys_params}
	\end{center}
\end{table*}

\begin{table}
	\begin{center}
		\begin{tabular}{|cc|} \hline
			Tuple of Infectious States' Fractions & Value \\ \hline
			$(S, I, R, P)$\textsubscript{PI} & $(0.46, 0.06, 0.46, 0.02)$ \\ 
			$(S, I, R, P)$\textsubscript{PI-free} & $(0.79, 0.005, 0.205, 0)$ \\ \hline
		\end{tabular}
		\caption{The various infectious states for the populations of every farm upon initialisation. The fraction with the underscript `PI' denotes farms destined to have a non-zero PI percentage, while that with `PI-free' farms free from any PI animals. The PI population tuple fractions were randomly allocated to 2\% of the input farms upon initialisation \cite{GET18}.}\label{tab:farm_infect}
	\end{center}
\end{table}


\begin{table*}
	\begin{center}
    \begin{scriptsize}
			\begin{tabular}{ccc} \hline
				Parameter & Value & Description \\ \hline
				Small Farm Margin & 10 & Population of a farm below which a farm is classified as of small farm type \\ 
				Farm Size\textsubscript{min} & 10 & Farm population above which farms are retained in the input \\
				Farm Size\textsubscript{max} & 10,000 & Farm population below which (inclusive) farms are retained in the input \\
				Slaughterhouses & 1 & Number of additional farms of the slaughterhouse type in the simulation \\
				Wells & 1 & Number of additional farms of the well type in the simulation \\
				Slaught. capacity & 10,000 & Number of animals a slaughterhouse can accept in a single call from a farm \\
				Well yield & 10,000 & Number of animals a well can introduce in a single call to a farm \\
				Well yield in TI & 2\% & The percentage of TI animals in every farm call from the well \\ 
				Well yield in PI & 2\% & The percentage of PI animals in every call from the well \\ 
				Threshold buy & 0 & The global threshold above which farms can buy animals \\
				Threshold sell & 0 & The global threshold above which farms can sell animals \\ 
				Replacement & 0.0054 & Relevant term in equation \eqref{eq:replacement} (on a herd's annual rejuvenation) \\
				Infectious margin prob. & 2\% & Selection threshold below which the population of a randomly initialised farm has a PI fraction \\ \hline
			\end{tabular}
    \end{scriptsize}
		\caption{Farm related parameters according to \cite{GET15a} and \cite{GET18}. All the parameters can be controlled from the input ini file.}\label{tab:farm_params}
	\end{center}
\end{table*}


\begin{table}
	\begin{center}
		\begin{tabular}{|c|c|} \hline
			Death & Survival \\ \hline
			2\%   & 98\%     \\ \hline
		\end{tabular}
		\caption{Effect probabilities (complementary) of BVD on calf in case of infection. After \cite{GET15a}.}\label{tab:calf_TI_results}
	\end{center}
\end{table}

\begin{table}
	\begin{center}
		\begin{tabular}{|c|c|c|c|c|} \hline
			Pregnancy periods & PI & Abortion & Malformation & R \\ \hline
			1\textsuperscript{st}: [0-70) days & 90 \% & 100\% & 0\% & 0\% \\ 
			2\textsuperscript{nd}: [70-120) days & 45\%  & 60\% & 75\% & 100\% \\ 
			3\textsuperscript{d}:  [120-180) days & 0\%   & 20\% & 45\% & 100\%  \\ 
			4\textsuperscript{th}: [180-max) days & 0\%   & 5\%  & 20\% & 100\% \\ \hline
		\end{tabular}
		\caption{Calf outcome mass (cumulative) probabilities in case of infection (horizontal) during pregnancy, max=[280-292). After \cite{GET15}.}\label{tab:embryo_outcome_from_TI}
	\end{center}
\end{table}

\begin{table}
	\begin{center}
		\begin{tabular}{ |c|c|c|c| }
			\hline
			\multicolumn{1}{|c|}{Insemination No} & \multicolumn{1}{|c|}{Heifers} & \multicolumn{1}{|c|}{Dams} \\
			\hline
			1 & 90.48\% & 67.03\%  \\
			2 & 99.53\% & 93.84\% \\
			3 & 99.98\% & 99.2\% \\
			4 & 100\% & 99.92\% \\ \hline 
		\end{tabular}
		\caption{Mass (cumulative) probabilities for a successful insemination of heifers or cows out of four total inseminations. Note that only cows have a 0.08\% probability to be declared infertile. After \cite{GET15a}.}\label{tab:insem_results}
	\end{center}
\end{table}

\begin{table}
	\begin{center}
		\begin{tabular}{|c|c|} \hline
			Stage & Probability	\\ \hline
			1\textsuperscript{st} [0-60) days & 7\% \\ 
			2\textsuperscript{nd} [60-90) days & 9\% \\
			3\textsuperscript{d} [90-120) days & 10\% \\
			4\textsuperscript{th} [120-210) days & 12\% \\ \hline
		\end{tabular}
		\caption{Abortion mass (cumulative) probabilities dependent on the stage of pregnancy. After \cite{GET15a}.}\label{tab:gen_abort_stage}
	\end{center}
\end{table}

\begin{table}
	\begin{center}
		\begin{tabular}{|c|c|} \hline
			Age & Probability \\ \hline
			1\textsuperscript{st} year [0-365) days & 50\% \\
			2\textsuperscript{nd} year [365-730) days & 67\% \\
			3\textsuperscript{d} year [730-1095) days & 72\% \\
			4\textsuperscript{th} year [1,095-1,460) days & 73,5\% \\ 
			Further years [1,460-3,650) days & 100\% \\ \hline
			
		\end{tabular}
		\caption{PI death mass (cumulative) probabilities dependent on age. After \cite{GET15a}.}\label{tab:PI_death_age}
	\end{center}
\end{table}

\begin{table}
	\begin{center}
		\begin{tabular}{|c||c|c|} \hline
			Random process & Max & Min  \\ \hline
			PD & 292 & 280  \\
			TI $\Delta t$ CD & 7 days & 0 days \\ 
			Female & 50 & 0 \\ \hline
		\end{tabular}
		\caption{Uniform distributions used in the simulation: Pregnancy duration (PD), calf time of death by infection (TI $\Delta t$ CD), sex determination (female). After \cite{GET15a}.}\label{tab:unif_distrs}
	\end{center}
\end{table}

\begin{table}
	\begin{center}
		\begin{tabular}{|c||c|c|c|} \hline
			Random process & Min & Max & Mode \\ \hline
			IAAD & 0 & 3,000 & 200 \\
			TFT	& 4 days & 30 days & 11 days \\
			IIT	& 18 days & 24 days & 21 days \\ 
			FIA & 480 days & 600 days & 540 days \\
			CA No & 3 & 5 & 4 \\
			DI & 4 days   & 8 days & 7 days \\
			TR & 42 days & 115 days & 90 days \\
			MAD & 180 days & 270 days & 210 days \\
			MLE\textsuperscript{1st} & 0 days & 30 days & 10 days \\
			MLE\textsuperscript{2nd} & 170 days & 250 days & 200 days \\
			MLE\textsuperscript{3d} & 450 days & 750 days & 640 days \\ \hline
		\end{tabular}
		\caption{Triangular distributions used in the simulation: Initial Animals' Age Distribution (IAAD), time of first test (TFT), inter-insemination time (IIT), first insemination age (FIA), number of calvings (CA No), duration of infection (DI), time of rest (TR), maternal antibody duration (MAD), male life expectancy (MLE). After \cite{GET15a}.}\label{tab:triang_distrs}
	\end{center}
\end{table}

\clearpage

\section{Control Strategies}

To emulate the network's behavior and to predict the effect of different counter measures on the PI prevalence we formulate a simulation plan of intervention strategies \cite{TIN12a,DAM15,CON15,WER17,THU17,IOT17}. This plan consists of 13 different scenarios, one baseline where the system is allowed to evolve freely without any intervention strategy and 12, each of which contains some sort of mitigation strategy for BVD, applied at different times, after the free system's dynamics (baseline) have settled to an equilibrium. The particular ordering of the strategies in each scenario and consequently the number of scenarios were dictated by the needs of a cost-benefit analysis performed by collaborators of the FLI \cite{GET18} and similar to a recent work done for the federal state of Styria in Austria \cite{MAR18a}.

\begin{table*}[h!]
	\begin{center}
		\begin{tabular}{|c|l|} \hline
			Strategy & Description \\ \hline
			1 		 & No control (baseline) \\
			2 		 & Old regulation \\
			3 		 & New regulation \\
			4 		 & New regulation and vaccination \\
			5a 		 & New regulation and YCW with a semesterly frequency \\
			5b 		 & New regulation and YCW with an annual frequency \\
			6a 		 & New regulation, vaccination and YCW with a semesterly frequency \\
			6b 		 & New regulation, vaccination and YCW with an annual frequency \\
			7 		 & YCW with a semesterly frequency \\
			8 		 & Vaccination \\
			9 		 & Vaccination and YCW with a semesterly frequency \\ \hline			
		\end{tabular}
		\caption{The different strategies comprising the scenarios presented in Tab.~\ref{tab:Scens}. YCW is the young calf window protocol. See the testing description in section \ref{sec:Submodels}.}\label{tab:Strats}
	\end{center}
\end{table*}

We first list the different control strategies and then their application protocol manifested in the distinct scenarios in Tab.~\ref{tab:Strats}. Similarly we summarize the scenarios with the strategies being applied at targeted times in Tab.~\ref{tab:Scens}. One time 
step should correspond to a day. The total running time of our simulations spanned 20,000 days or roughly 55 years, in which we 
distributed the different strategies. The reasoning has been always to start from a `no control' state in equilibrium and apply 
some control strategy on the system once its dynamics have settled on a fixed point or have reached a state of slow variation. 

\begin{table}
	\begin{center}
		\begin{tabular}{|ccccccc|} \hline
			\multirow{2}{*}{Scenarios}
			& \multicolumn{6}{c|}{Timeline (in days)} \\ \cline{2-7} 
			& 0 & 10,000 & 12,006 & 12,373 & 12,738 & 20,000 \\ \hline
			1 & \multicolumn{6}{c|}{\cellcolor[HTML]{6195C9}STR 1} \\ 
			2 &  \multirow{12}{*}{\cellcolor[HTML]{6195C9}STR 1} & \multicolumn{5}{c|}{\cellcolor{gray!20}STR 2} \\
			3 & \cellcolor[HTML]{6195C9} &  \multirow{11}{*}{\cellcolor{gray!20}STR 2} & \multicolumn{4}{c|}{\cellcolor{LightCyan}STR 3} \\
			4 & \cellcolor[HTML]{6195C9} & \cellcolor{gray!20} & \multicolumn{2}{c}{\cellcolor{LightCyan}} & \multicolumn{2}{c|}{\cellcolor{gray!20}STR 1} \\ 
			5 & \cellcolor[HTML]{6195C9} & \cellcolor{gray!20} & \cellcolor{LightCyan} & \multicolumn{3}{c|}{\cellcolor[HTML]{6195C9}STR 4} \\
			6 & \cellcolor[HTML]{6195C9} & \cellcolor{gray!20} & \cellcolor{LightCyan} & \multicolumn{3}{c|}{\cellcolor{gray!20}STR 5a} \\
			7 & \cellcolor[HTML]{6195C9} & \cellcolor{gray!20} & \cellcolor{LightCyan} & \multicolumn{3}{c|}{\cellcolor[HTML]{6195C9}STR 5b} \\
			8 & \cellcolor[HTML]{6195C9} & \cellcolor{gray!20} & \cellcolor{LightCyan} & \multicolumn{3}{c|}{\cellcolor{gray!20}STR 6a} \\
			9 & \cellcolor[HTML]{6195C9} & \cellcolor{gray!20} & \cellcolor{LightCyan} & \multicolumn{3}{c|}{\cellcolor[HTML]{6195C9}STR 6b} \\
			10 & \cellcolor[HTML]{6195C9} & \cellcolor{gray!20} & \cellcolor{LightCyan} & \multicolumn{3}{c|}{\cellcolor{gray!20}STR 7} \\
			11 & \cellcolor[HTML]{6195C9} & \cellcolor{gray!20} & \cellcolor{LightCyan} & \multicolumn{3}{c|}{\cellcolor[HTML]{6195C9}STR 8} \\
			12 & \cellcolor[HTML]{6195C9} & \cellcolor{gray!20} & \cellcolor{LightCyan} & \cellcolor{gray!20} STR4 & \multicolumn{2}{c|}{\cellcolor{green} STR8} \\
			13 & \cellcolor[HTML]{6195C9} & \cellcolor{gray!20} & \cellcolor{LightCyan} & \cellcolor{green} STR4 & \multicolumn{2}{c|}{\cellcolor{gray!20} STR9} \\ \hline
		\end{tabular}
		\caption{The scenario scheduling plan. STR stands for `strategy' as outlined in Tab.~\ref{tab:Strats}. A color block denotes the effect of the same strategy throughout the different starting days.}\label{tab:Scens}
	\end{center}
\end{table}

According to Tab.~\ref{tab:Scens}, the 10,000\textsuperscript{th} step of any of the mitigation strategies signifies the start of the old regulation's effect (2011).  The nationwide PI prevalence before 2011 is not known due to lack of unified records. Hence, we restrict ourselves to the 10,000\textsuperscript{th} time step to mark the initiations of the intervention strategies, as a large enough time where the population dynamics appear to have stochastically settled to a fixed point.
In other words, our mapping of the 10,000\textsuperscript{th} time step with the commencement of the year 2011 (enforcement of the old regulation) is only speculative and cannot be precisely determined exactly due to the prior non-nationwide, but regional intervention strategy programs.

\section{Summary}
In this article we have introduced -- on biological, data, policy and agricultural grounds -- a stochastic, event-driven agent-based model to emulate the current situation of BVD in Germany via within-farm contact mechanisms and through animal movement contacts. We have furthermore provided a thorough description following a specific protocol (ODD). The proposed model includes a number of mitigation strategies of interest to a cost-benefit analysis. 

\begin{acknowledgements}
JB and PH acknowledge the support by Deutsche Forschungsgemeinschaft (DFG) in the framework of Collaborative Research Center 910 and German Academic Exchange Serivce (DAAD) via the PPP/PROCOPE program. The authors acknowledge helpful discussions with A. Koher and H.~H.~K. Lentz.
\end{acknowledgements}


\begin{thebibliography}{10}
\expandafter\ifx\csname url\endcsname\relax
  \def\url#1{{\tt #1}}\fi
\expandafter\ifx\csname urlprefix\endcsname\relax\def\urlprefix{URL }\fi

\bibitem{GRI10}
V.~Grimm, U.~Berger, D.~L. DeAngelis, J.~G. Polhill, J.~Giske, and S.~F.
  Railsback: {\em The {ODD} protocol: a review and first update\/}, Ecol.
  Model. {\bf 221}, 2760 (2010).

\bibitem{THU17}
H.~H. Thulke, M.~Lange, J.~A. Tratalos, T.~A. Clegg, G.~McGrath, L.~O'Grady,
  P.~O'Sullivan, M.~L. Doherty, D.~A. Graham, and S.~J. More: {\em Eradicating
  bvd, reviewing irish programme data and model predictions to support
  prospective decision making\/}, Prev. Vet. Med. {\bf 150}, 151 (2017).

\bibitem{FIS13}
G.~Fishman: {\em Discrete-event simulation: modeling, programming, and
  analysis\/} (Springer Science \& Business Media, 2013).

\bibitem{SKI98}
S.~S. Skiena: {\em The algorithm design manual: Text\/}, vol.~1 (Springer,
  1998).

\bibitem{ALL08}
L.~J.~S. Allen, F.~Brauer, P.~Van~den Driessche, and J.~Wu: {\em Mathematical
  epidemiology\/}, vol. 1945 (Springer, 2008).

\bibitem{BRA12}
F.~Brauer and C.~Castillo-Chavez: {\em Mathematical models in population
  biology and epidemiology\/}, vol.~40 (Springer, 2012).

\bibitem{GET15a}
J.~Gethmann: {\em Programm zur {S}imulation der {BVD}-{S}anierung\/} (2015),
  personally prepared doc file.

\bibitem{GET18}
{\em Personal communication with {D}r. {J}. {G}ethmann, from the
  {F}riedrich-{L}oeffler {I}nstitute\/}.

\bibitem{GNU18}
M.~Galassi {\em et~al.\/}: {\em GNU Scientific Library Reference Manual (3rd
  Ed.), ISBN 0954612078\/} (2018), {http://www.gnu.org/software/gsl/}.

\bibitem{KEE05}
K.~T.~D. Eames: {\em Networks and epidemic models\/}, Journal of The Royal
  Society Interface {\bf 2}, 295 (2005).

\bibitem{VIE04}
A.~F. Viet, C.~Fourichon, H.~Seegers, C.~Jacob, and C.~Guihenneuc-Jouyaux: {\em
  {A model of the spread of the bovine viral-diarrhoea virus within a dairy
  herd}\/}, Prev. Vet. Med. {\bf 63}, 211 (2004).

\bibitem{KAM03}
N.~G. van Kampen: {\em Stochastic Processes in Physics and Chemistry\/}
  (North-Holland, Amsterdam, 2003).

\bibitem{CON15}
F.~J. Conraths and J.~Gethmann: {\em Epidemiologische untersuchungen in
  tierpopulationen\/}, Tech. rep., FLI, Greifswald - Insel Riems (2015).

\bibitem{KRI06}
K.~Krishnamoorthy: {\em Handbook of statistical distributions with
  applications\/} (Chapman and Hall/CRC, 2006).

\bibitem{GET15}
J.~Gethmann, T.~Homeier, M.~Holsteg, H.~Schirrmeier, M.~Sa{\ss}erath,
  B.~Hoffmann, M.~Beer, and F.~J. Conraths: {\em {BVD-2 outbreak leads to high
  losses in cattle farms in Western Germany}\/}, Heliyon {\bf 1}, e00019
  (2015).

\bibitem{EZA07}
P.~Ezanno, C.~Fourichon, A.~F. Viet, and H.~Seegers: {\em Sensitivity analysis
  to identify key-parameters in modelling the spread of bovine viral diarrhoea
  virus in a dairy herd\/}, Prev. Vet. Med. {\bf 80}, 49 (2007).

\bibitem{BIO16}
L.~Bioglio, M.~G{\'e}nois, C.~L. Vestergaard, C.~Poletto, A.~Barrat, and
  V.~Colizza: {\em Recalibrating disease parameters for increasing realism in
  modeling epidemics in closed settings\/}, BMC Infect. Dis. {\bf 16}, 676
  (2016).

\bibitem{DAM15}
A.~Damman, A.~F. Viet, S.~Arnoux, M.~C. Guerrier-Chatellet, E.~Petit, and
  P.~Ezanno: {\em Modelling the spread of bovine viral diarrhea virus ({BVDV})
  in a beef cattle herd and its impact on herd productivity\/}, Vet. Res. {\bf
  46}, 12 (2015).

\bibitem{LEG08}
{\em Verordnung zum {S}chutz der {R}inder vor einer {I}nfektion mit dem
  {B}ovinen {V}irusdiarrhoe-{V}irus, decreed by the federal ministry of food
  and agriculture\/} (2008).

\bibitem{LEG08a}
{\em Zweite {V}erordnung zur \"{A}nderung der {BVDV}-{V}erordnung\/} (2016).

\bibitem{TIN12a}
M.~R. Tinsley, F.~I. Lewis, and F.~Br{\"u}lisauer: {\em Network modeling of
  {BVD} transmission\/}, Vet. Res. {\bf 43} (2012).

\bibitem{WER17}
K.~Wernike, J.~Gethmann, H.~Schirrmeier, R.~Schr{\"o}der, F.~J. Conraths, and
  M.~Beer: {\em Six years (2011--2016) of mandatory nationwide bovine viral
  diarrhea control in {G}ermany -- a success story\/}, Pathogens {\bf 6}, 50
  (2017).

\bibitem{IOT17}
B.~Iotti, E.~Valdano, L.~Savini, L.~Candeloro, A.~Giovannini, S.~Rosati,
  V.~Colizza, and M.~Giacobini: {\em Farm productive realities and the dynamics
  of bovine viral diarrhea (bvd) transmission\/}, bioRxiv p. 230045 (2017).

\bibitem{MAR18a}
T.~Marschik, W.~Obritzhauser, P.~Wagner, V.~Richter, M.~Mayerhofer,
  C.~Egger-Danner, A.~K{\"a}sbohrer, and B.~Pinior: {\em A cost-benefit
  analysis and the potential trade effects of the bovine viral diarrhoea
  eradication programme in {S}tyria, {A}ustria\/}, The Vet. J. {\bf 231}, 19
  (2018).

\end{thebibliography}
\end{document}